\documentclass[%
 aip,
 amsmath,amssymb,
 reprint,%
]{revtex4-1}

\usepackage{graphicx}
\usepackage{dcolumn}
\usepackage{bm}
 \usepackage[english]{babel}

\usepackage[utf8]{inputenc}
\usepackage[T1]{fontenc}
\usepackage{lineno}
\usepackage{mathptmx}
\usepackage{etoolbox}
\usepackage{amsmath}
\usepackage{braket}
\usepackage{xcolor}
\usepackage{float}
\makeatletter
\def\@email#1#2{%
 \endgroup
 \patchcmd{\titleblock@produce}
  {\frontmatter@RRAPformat}
  {\frontmatter@RRAPformat{\produce@RRAP{*#1\href{mailto:#2}{#2}}}\frontmatter@RRAPformat}
  {}{}
}%
\makeatother
\begin{document}

\newcommand{\xhat}{{\bf \hat{x}}}
\newcommand{\yhat}{{\bf \hat{y}}}
\newcommand{\zhat}{{\bf \hat{z}}}
\newcommand{\ehat}{{\bf \hat{e}}}
\newcommand{\BE}{{\bf E}}
\newcommand{\ayan}[1]{\textcolor{black}{ #1}}
\newcommand{\GKS}[1]{\textcolor{black}{ #1}}

\preprint{AIP/123-QED}

\title {Beam-splitter-free, high-rate quantum key distribution inspired by intrinsic quantum mechanical spatial randomness of entangled photons}
\author{Ayan Kumar Nai}
\altaffiliation[Author to whom correspondence should be addressed: ayankrnai@gmail.com]{}
\affiliation{Photonic Sciences Lab., Physical Research Laboratory, Ahmedabad 380009, Gujarat, India}
\affiliation{Indian Institute of Technology Gandhinagar, Palaj, Gandhinagar 382055, Gujarat, India}

%
\author{Gopal Prasad Sahu}%
\affiliation{Indian Institute of Technology Dharwad, Dharwad  580011, Karnataka, India}
\author{Rutuj Gharate}%
\affiliation{Photonic Sciences Lab., Physical Research Laboratory, Ahmedabad 380009, Gujarat, India}

\author{C. M. Chandrashekar}%
\affiliation{Quantum Optics $\&$ Quantum Information, Department of Electronic and Systems Engineering, Indian Institute of Science, Bengaluru 560012, India} 

\author{G. K. Samanta}
\affiliation{Photonic Sciences Lab., Physical Research Laboratory, Ahmedabad 380009, Gujarat, India}


\begin{abstract}
Quantum key distribution (QKD) using entangled photon sources (EPS) is a cornerstone of secure communication. Despite rapid advances in QKD, conventional protocols still employ beam splitters (BSs) for passive random basis selection. However, BSs intrinsically suffer from photon loss, imperfect splitting ratios, and polarization dependence, limiting the key rate, increasing the quantum bit error rate (QBER), and constraining scalability, particularly over long distances. By contrast, EPSs based on spontaneous parametric down-conversion (SPDC) intrinsically exhibit quantum randomness in spatial and spectral degrees of freedom, offering a natural replacement for BS-based basis selection. Here, we demonstrate a proof-of-concept QKD scheme that exploits the intrinsic spatial randomness of SPDC without employing beam splitters. The annular SPDC emission ring is divided into four spatial sections, effectively generating two independent EPSs whose photon pairs are distributed to Alice and Bob. \GKS{Crucially, the measurement basis is not predetermined but is assigned after photon detection by exploiting intrinsic detector timing jitter, thereby concealing the basis information from a potential eavesdropper. This post-detection basis assignment emulates stochastic basis choice while avoiding BS-induced losses and bias.}
Experimentally, our scheme achieves a 6.4-fold enhancement in sifted key rate, a consistently reduced QBER, and a near-ideal encoding balance between linear and rectilinear bases. Furthermore, the need for four spatial channels can be avoided by employing wavelength demultiplexing to generate two EPSs at distinct wavelength pairs. Harnessing intrinsic spatial/spectral randomness thus enables robust, bias-free, high-rate, and low-QBER QKD, offering a scalable pathway for next-generation quantum networks.

\end{abstract}

\maketitle

\section{Introduction}
Classical communication underpins the modern digital age, with data security relying on the computational complexity of breaking encryption schemes \cite{gisin2002quantum, shor1999polynomial}. However, the rapid progress in quantum computing, with scalable high-quality qubits on the horizon, poses a profound threat to this security. In particular, Shor’s algorithm \cite{shor1999polynomial, ekert1996quantum, beckman1996efficient} exposes the vulnerability of widely used classical encryption methods to quantum attacks, underscoring the urgent need for more resilient solutions.

Quantum key distribution (QKD), on the other hand, has emerged as a robust alternative for secure key exchange \cite{bennett2014quantum, ekert1991quantum, gisin2010quantum, yin2017satellite}. QKD guarantees security even against attempts by an adversary to intercept the key \cite{shor2000simple, lucamarini2018, gottesman2004security}, thereby remaining unaffected by advances in computational power, including those offered by quantum computers. As a result, the QKD protocols have progressed from laboratory demonstrations to real-world deployment through various protocols, including BB84 \cite{gisin2002quantum, bennett2014quantum}, SARG04 \cite{scarani2004quantum}, COW \cite{stucki2009high}, and E91 \cite{ekert1991quantum}. Among these, BB84, the prepare and measure protocol based on weak coherent pulses, has emerged as the most practical choice for implementation, owing to its simplicity as well as the availability of composable security proofs \cite{shor2000simple, pirandola2020}, despite known vulnerabilities to specific attacks \cite{lucamarini2012device, Huttner:95}. 
In parallel, entanglement-based QKD (EBQKD) protocols \cite{ekert1991quantum, waks2002security, brassard2000limitations} have evolved, offering enhanced security owing to two fundamental quantum principles, the no-cloning theorem \cite{Scarani:05} and entanglement monogamy \cite{peev2009, coffman2000distributed, xu2020secure}. While EBQKD has been demonstrated over long free-space links by deploying entangled photon sources on satellites to enable key exchange between distant stations \cite{yin2017satellite, liao2018satellite, villar2020entanglement}, its security, verified through entanglement measurements such as Bell inequality violations \cite{ekert1991quantum}, ensures robustness against a wide spectrum of sophisticated attacks.
Alternatively, the BBM92 protocol \cite{bennett1992quantum} enables secure key distribution without the need for continuous monitoring of Bell inequalities, offering higher key rates and reduced resource requirements provided that the shared state preserves maximal entanglement. In this framework, security is ensured through post-processed quantum bit error rate (QBER) evaluation combined with occasional Bell inequality verification, allowing device-independent operation \cite{masanes2011secure} without relying on trusted nodes \cite{waks2002security, scherer2011long}. This positions EBQKD as a strong candidate for future quantum networks. Nonetheless, practical implementations face inherent challenges, including low key rates originating from the limited brightness of spontaneous parametric down-conversion (SPDC) sources \cite{harris1967observation}, and imperfections in 50:50 beam splitters used for random basis (Z and X) selection, thus introducing encoding bias and increased QBER \cite{mishra2022bbm92}. Although brightness can be improved through the use of high figure-of-merit nonlinear materials and optimized system configurations \cite{steinlechner2014efficient, jabir2017robust}, the use of beam splitters remains a critical bottleneck. In particular, beam splitters not only suffer from practical deviations from the ideal 50:50 ratio, but also impose a fundamental 50$\%$ loss in conventional schemes \cite{pirandola2020, li2022security}, thereby limiting achievable key rates. It is therefore crucial to explore new experimental strategies that achieve high key rates and low QBER while eliminating the dependence on beam splitters. 

Here, we demonstrate a high-performance implementation of the BBM92 protocol that exploits the intrinsic spatial randomness of SPDC photons. \GKS{Using a type-0 phase-matched PPKTP crystal embedded in a polarization Sagnac interferometer, we generate polarization-entangled Bell states at diametrically opposite positions on the annular spatial emission of the photon pairs. The use of intrinsic quantum randomness of correlated photon pairs emitted at opposite points on the SPDC ring \cite{nai2025beam}, together with post-detection basis assignment enabled by intrinsic detector timing jitter, our approach eliminates the need for beam splitters in random basis selection and thereby avoids beam-splitter-induced loss and bias.} Experimentally, we observe a 6.4-fold enhancement in the sifted key rate and a consistently reduced quantum bit error rate (QBER) compared with conventional beam-splitter-based BBM92 implementations. Furthermore, the scheme achieves a near-ideal balance between the $Z$ (H/V) and $X$ (D/A) bases, enabling reliable and low-QBER key generation.


\section{Experiment}
\label{experiment}
\begin{figure*}[ht]
    \centering
    \includegraphics[width=\linewidth]{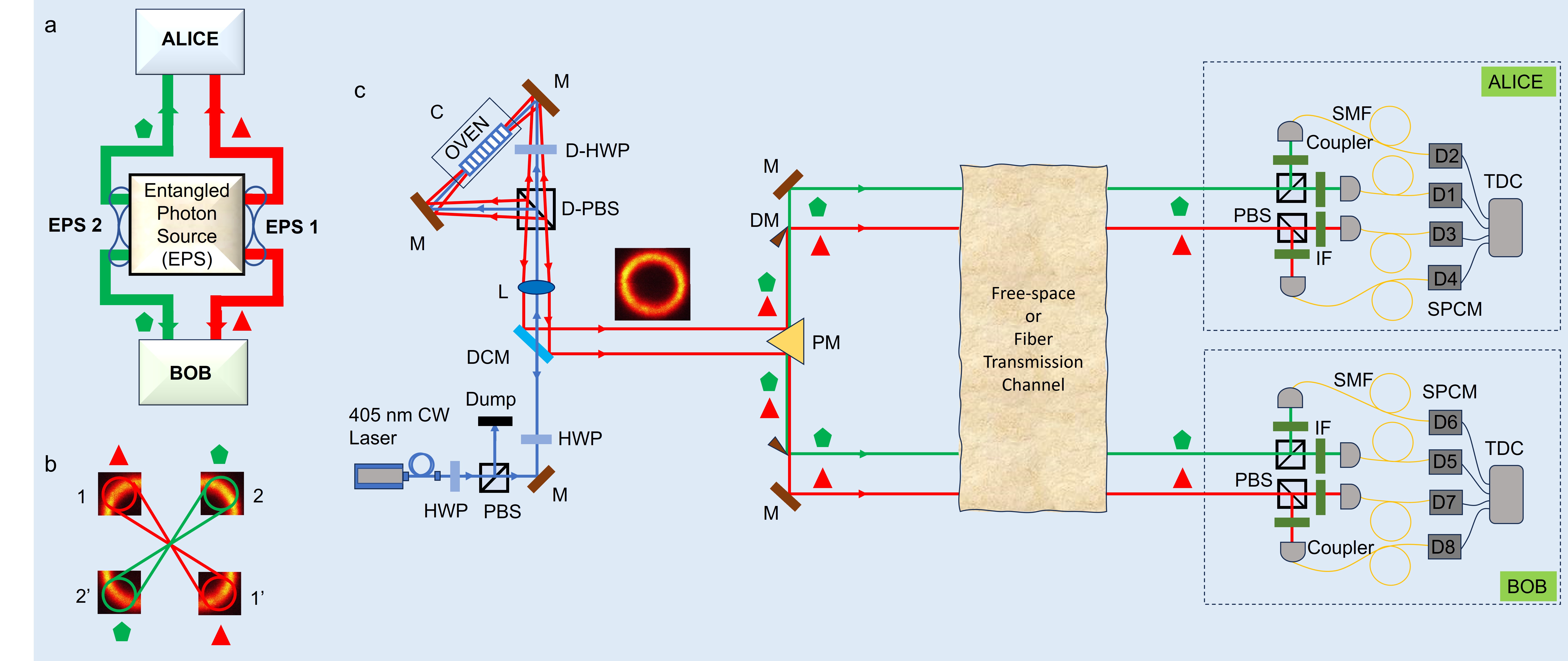}
    \caption{\textbf{Experimental setup and network architecture.} \textbf{a}, Conceptual representation of the present BBM92 scheme. \textbf{b}, Division of the spatial distribution of pair photons to form two identical entangled photon sources. \textbf{c}, Schematic of the experimental setup. Laser: 405 nm cw diode laser; HWP: half-wave plate; PBS: polarizing beam splitter; M: mirrors; DCM: dichroic mirror; L: plano-convex lenses; D-PBS: dual-wavelength (405 nm and 810 nm) PBS; D-HWP: dual-wavelength HWP; C: PPKTP crystal in an oven for photon-pair generation; PM: prism-shaped gold-coated mirror; DM: D-shaped mirrors; IF: 3 nm bandwidth interference filter; Coupler: collimator system for fiber coupling; SMF: single-mode fiber; SPCM: single-photon counting module; TDC: time-to-digital converter. 
    }
    \label{Figure1}
\end{figure*}

The concept of the spatial randomness–based BBM92 scheme is illustrated in Fig. \ref{Figure1}(a). The entangled photon source (EPS) can generate any of the four Bell states, $\ket{\Phi^{\pm}}$ and $\ket{\Psi^{\pm}}$. By applying a spatial-division scheme \cite{nai2024device} to the annular ring distribution of the single EPS, constructed from a common set of physical resources including pump laser, nonlinear crystal, and interferometric architecture, we obtain two (see Fig. \ref{Figure1}(b)) effective time-multiplexed EPSs (EPS 1 and EPS 2). The generation of pair photons from EPS 1 and EPS 2 is intrinsically quantum-random in time and unbiased. These pair photons are distributed between two users, Alice and Bob. Logical bits are encoded as the conventional schemes, with “0” and “1” corresponding to detections in the $H/D$ and $V/A$ polarizations, respectively. 

The schematic of the experimental setup is shown in Fig. \ref{Figure1}(c). A continuous-wave, single-frequency diode laser delivering 40 mW of output power at a central wavelength of 405 nm, with a linewidth of approximately 20 MHz, is used as the pump source. A polarizing beam splitter (PBS) cube, in combination with a half-wave plate (HWP), is used to control the incident pump power. A second HWP, oriented with its fast axis at 22.5$^\circ$ relative to the vertical, rotates the horizontally polarized pump beam into a diagonal state. The pump is then focused to a waist diameter of $\sim$40 µm at the center of the nonlinear crystal using a plano-convex lens (L) with focal length $f=150$ mm. The nonlinear medium is a 10-mm-long periodically poled potassium titanyl phosphate (PPKTP) crystal (SLF Svenska Laserfabriken AB) with a $2\times 1$ mm$^2$ in aperture and a single grating period $\Lambda=3.425$ $\mu$m. The crystal, housed in a temperature-controlled oven (stability $\pm$0.1$^\circ$C, tunable up to 200$^\circ$C), is positioned at the center of a polarization Sagnac interferometer formed by a dual-wavelength PBS (D-PBS, designed for 405 and 810 nm) and two plane mirrors (M). A dual-wavelength half-wave plate (D-HWP) placed inside the Sagnac interferometer, with its fast axis at 45$^\circ$, flips the polarization between horizontal and vertical. Under these conditions, the crystal produces non-collinear, degenerate, type-0 ($e \rightarrow e + e$) pair photons at 810 nm via quasi-phase-matched spontaneous parametric down-conversion (SPDC). As described in earlier work \cite{jabir2017robust, Singh:21}, the clockwise and counterclockwise pump components inside the Sagnac interferometer generate pair photons with orthogonal polarizations, which combine at the D-PBS to produce the maximally entangled state, $\ket{\Phi^{+}} = \frac{1}{\sqrt{2}} (\ket{HH} + \ket{VV})$.
The annular distribution of the down-converted photons is collimated by the lens (L) and separated from the pump by a dichroic mirror (DCM). The SPDC ring is divided into two halves using a prism-shaped gold-coated mirror (PM), and each half is further subdivided into two sections by D-shaped mirrors (DM), yielding four spatial regions denoted 1, 2, 1$'$, and 2$'$. For visualization, pair photons from regions (1, 1$'$) and (2, 2$'$) forming EPS 1 and EPS 2 are labeled as red triangles and green pentagons, respectively. The spatial sections (1, 2) and (1$'$, 2$'$) of the SPDC ring are distributed between Alice and Bob, respectively. \GKS{The pair photons of each sections are projected in two orthogonal polarizations states using the PBS. } The photons of the output ports of the PBSs are spectrally filtered by $\sim$3 nm interference filters (IF), coupled into single-mode fibers (SMFs), and measured by single-photon counting modules (SPCMs)  connected to the time-to-digital converter (TDC) interfaced with a computer for coincidence detection and data processing. \GKS{Both Alice and Bob have four SPCMs, D1-4 and D5-8 respectively, connected to the TDC to record the detection time stamps of photons. Unless otherwise stated, the coincidence window is maintained at 1 ns throughout the manuscript. Alice uses detectors D1 and D2 to measure the transmitted and reflected outputs of PBS1 for photons from EPS1, while detectors D3 and D4 measure the corresponding outputs of PBS2 for photons from EPS2. Similarly, Bob uses detectors D5 and D6 to measure the transmitted and reflected outputs of PBS1 for photons from EPS1, while detectors D7 and D8 measure the corresponding outputs of PBS2 for photons from EPS2.}

As observed previously \cite{nai2025beam}, at any given instant EPS 1 and EPS 2 have equal probability of generating pair photons. \GKS{Therefore, all detectors have an equal probability of detecting photons, without the bias commonly observed in beam-splitter-based schemes due to the inherent imperfections of practical 50:50 beam splitters. Under ideal conditions, the photon count rates of all detectors are expected to be equal, provided that the detectors have identical detection efficiencies and all optical channels are equally coupled. Any imbalance in detector efficiency introduces similar artifacts in other QKD schemes as well. Thus, with proper calibration and alignment of the detectors, we can detect comparable photon count rates across all detectors in the present scheme eliminating the losses.}
\GKS{Now the simplest approach to achieve a high bit-rate QKD scheme is to assign fixed measurement bases, $Z$ or $X$, to photons originating from EPS1 and EPS2 for both Alice and Bob. However, such an arrangement introduces fundamental security vulnerabilities because the basis selection effectively occurs at the source, while EPS1 and EPS2 remain distinguishable either spatially or spectrally. As a result, an eavesdropper (Eve) may exploit side-channel information to infer the measurement basis without disturbing the quantum state. To mitigate this vulnerability, we assign the measurement basis only after photon detection by exploiting intrinsic detector timing jitter to conceal the basis information from Eve. Based on this principle, we propose two protocols, referred to as Protocol $\#$1 and Protocol $\#$2. In Protocol $\#$1, Alice and Bob record the detection timestamps ($t_A$ and $t_B$) from all detectors and publicly announce the timestamps together with the corresponding source label (EPS1 or EPS2), while keeping the detector identity private. They retain only those events originating from matched sources and further post-select coincident events satisfying $|t_A - t_B| \leq 1$ ns. For each coincident event, they compute $b = |t_A - t_B| \bmod 2$ and assign the measurement basis as $Z$ (H/V) for $b = 0$ and $X$ (D/A) for $b = 1$. Like the standard BBM92 scheme, here also the detections in the $Z$ and $X$ are encoded as logical bits “0” for the $H$ and $D$ and “1” for $V$ and $A$ polarization outputs. Since the basis assignment is derived jointly from coincident detection events, no basis mismatch occurs, leading to a fourfold increase in the sifted key rate compared to the conventional BB84 protocol. The security of this scheme, however, critically depends on strong assumptions, including trusted detectors, intrinsic and uncontrollable timing jitter, and the operational indistinguishability of the photon sources. In Protocol $\#$2, Alice and Bob independently record detection timestamps at their respective stations and determine the basis using the time difference between consecutive detections, defined as $b = |t_{i+1} - t_i| \bmod 2$, assigning $Z$ or $X$ bases for $b = 0$ or $1$, respectively. Similar to the BBM92 protocol, Alice and Bob publicly announce the timestamps, source information, and basis choices. Because the basis selection is performed independently, approximately $50\%$ of coincident events are discarded due to basis mismatch. Nevertheless, due to the use of two sources, this scheme still achieves a twofold improvement in sifted key generation rate relative to the standard protocol. Importantly, Protocol $\#$2 relies on weaker trust assumptions and therefore offers improved security robustness compared to Protocol $\#$1, but at the cost of reduced efficiency. Overall, Protocol $\#$1 prioritizes maximum key-rate enhancement under stringent detector-trust assumptions, whereas Protocol $\#$2 provides a more conservative and secure implementation with moderate efficiency gains.}

\GKS{Two entangled photon sources (EPS1 and EPS2) can also be derived from a single SPDC source through energy conservation, corresponding to correlated signal–idler wavelength pairs $(\lambda_1, \lambda_2)$ and $(\lambda_3, \lambda_4)$. The generation of photon pairs with different wavelength combinations that satisfy energy conservation is intrinsically random, analogous to the random spatial-mode distribution in SPDC. However, in the conventional BBM92 protocol, only one such wavelength pair (either EPS1 or EPS2) is utilized for QKD, while the remaining spectral components, corresponding to other correlated wavelength pairs within the SPDC bandwidth, are discarded. In contrast, the present approach exploits these otherwise unused spectral components to enable high–bit-rate implementation of the BBM92 protocol. This is achieved without increasing the number of quantum communication channels between the two users, requiring only the same two channels as in the standard BBM92 scheme.}


\GKS{Finally, the sifted key is subsequently processed using standard error correction (EC) and privacy amplification (PA) protocols to distill the final secure cryptographic key. It is important to emphasize that all experimental parameters reported here are derived directly from raw data without applying any corrections. While the implementation of such corrections could further improve the reported values, they do not hold practical significance for real-time applications of QKD systems. Hence, the uncorrected results presented here provide a more realistic performance assessment of the current protocol under operational conditions.}

\section{Results and discussions}
\label{RnD}
\begin{figure}[H]
    \centering
    \includegraphics[width=\linewidth]{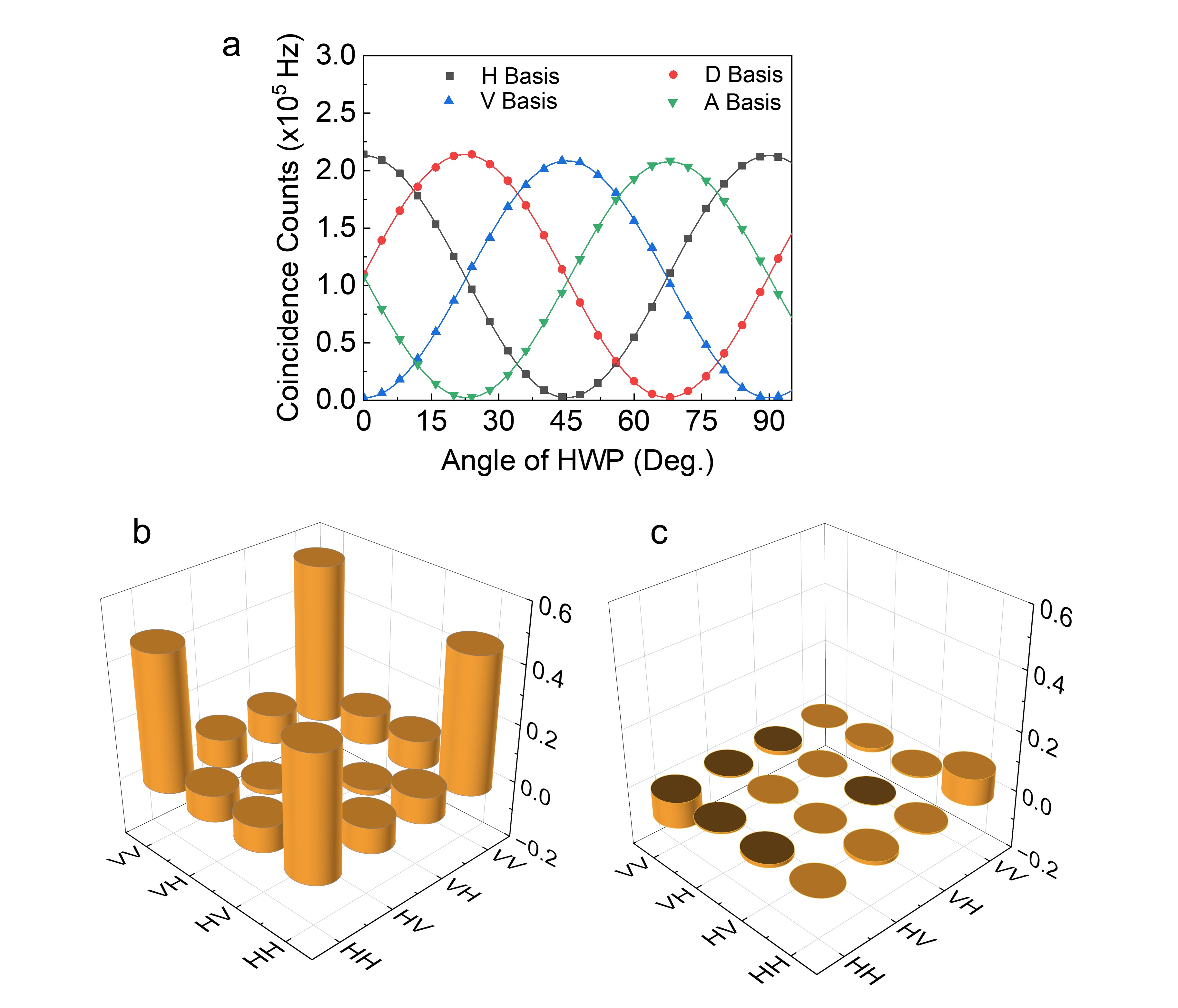}
    \caption{\textbf{Characterization of the entangled photon sources.} \textbf{a}, Quantum interference of spatially separated entangled photon sources measured in the horizontal (H, black dots), vertical (V, blue dots), diagonal (D, red dots), and anti-diagonal (A, green dots) polarization bases. Solid lines represent sinusoidal fits to the experimental data. Absolute values of \textbf{d}, real and \textbf{e}, imaginary parts of the reconstructed density matrix of the polarization-entangled Bell state $\ket{\phi^+}$.}
    \label{Figure2}
\end{figure}

We first characterized EPS 1 and EPS 2 in terms of their quantum performance parameters, including entanglement visibility, Bell parameter, and state fidelity. As both sources are derived from a single entangled photon source, they exhibit nearly identical characteristics. For clarity, we present the representative results of EPS 2 (see green paths in Fig. \ref{Figure1}(b)). The measurements, summarized in Fig. \ref{Figure2}, were performed at a fixed pump power of 3 mW and crystal temperature of T = 41.5$^\circ$C.

We recorded the coincidence counts between photons transmitted through the PBS of both Alice and Bob, with Alice having fast axis of the HWP fixed at $0^\circ$, $22.5^\circ$, $45^\circ$, and $67.5^\circ$, corresponding to the $H$ (black), $V$ (blue), $D$ (red), and $A$ (green) bases, respectively, while Bob continuously rotated the HWP from $0^\circ$ to $90^\circ$ in $4^\circ$ steps. The resulting interference fringes are shown in Fig. \ref{Figure2}(a). 
As shown in Fig. \ref{Figure2}(a), the coincidence counts exhibit the expected sinusoidal modulation, with measured fringe visibilities of 97.8$\%$, 97.8$\%$, 97.7$\%$, and 97.7$\%$ in the $H$, $V$, $D$, and $A$ bases, respectively—well above the quantum threshold of 71$\%$ \cite{jabir2017robust}, thereby confirming high-quality entanglement. Furthermore, by analyzing coincidence counts across 16 different combinations of HWP settings of Alice and Bob and applying the standard CHSH inequality formalism \cite{CHSH:69}, we obtained a Bell parameter of $S = 2.765 \pm 0.007$, clearly violating the classical bound by nearly 109 standard deviations. Further, we carried out quantum state tomography based on coincidence measurements across different polarization projections. The reconstructed real and imaginary parts of the two-qubit density matrix, shown in Fig. \ref{Figure2}(b) and \ref{Figure2}(c), respectively, confirm the generation of the maximally entangled Bell state $\ket{\Phi^{+}} = \tfrac{1}{\sqrt{2}}(\ket{HH} + \ket{VV})$ with a fidelity of 97$\%$. The source brightness of EPS 2 is estimated to be 0.04 MHz/mW/nm. We observed comparable performance parameters for EPS 1. These results establish the suitability of both EPS 1 and EPS 2 for the implementation of the QKD protocol.

\begin{figure}[H]
    \centering
    \includegraphics[width=\linewidth]{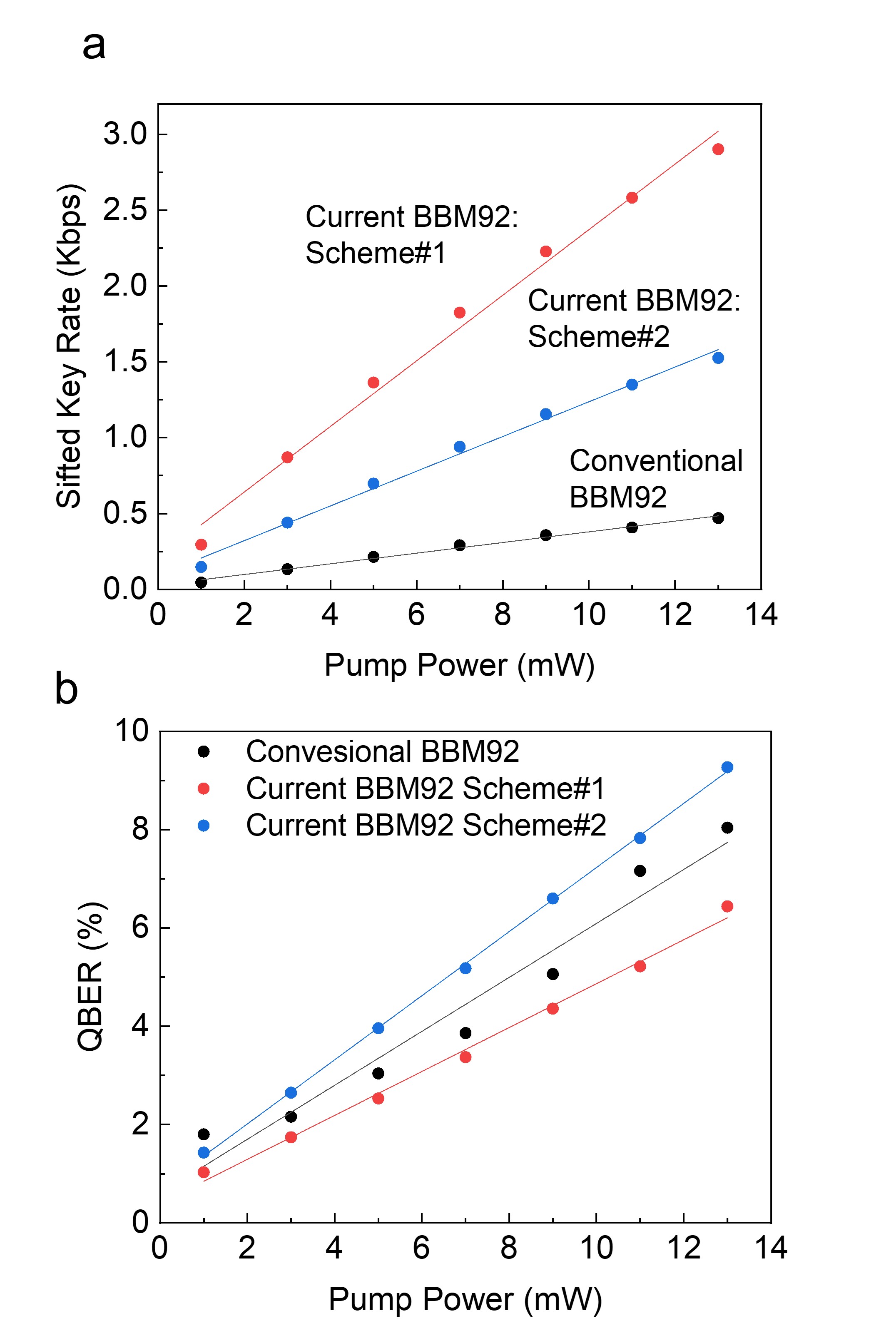}
    \caption{\textbf{Performance metrics of the QKD schemes.} (a) Sifted key rate and (b) quantum bit error rate (QBER) as functions of pump power for the \GKS{current BBM92 scheme with Protocol $\#$1 (red circles), Protocol $\#$2 (blue circles)} and the conventional BS-based BBM92 scheme (black circles). Solid lines represent linear fits to the experimental data.}
    \label{Figure3}
\end{figure}

With the quantum parameters of the entangled photon sources established, we implemented the BBM92 protocol and benchmarked its performance against the conventional beam-splitter–based approach. Keeping the crystal temperature fixed at $41.5^\circ$C, we measured the sifted key rate and QBER as functions of pump power. The results are shown in Fig. \ref{Figure3}. As expected, the sifted key rate (see Fig. \ref{Figure3}(a))  calculated by summing all the coincidence counts of same basis detection of two users (here, Alice and Bob), increases with pump power in both cases owing to the higher pair-generation rate. \GKS{In basis-randomization Protocol \#1 (red dots), the sifted key rate increases from 0.29 Mbps to 2.9 Mbps as the pump power is raised from 1 to 13 mW, corresponding to a slope of 0.22 Mbps/mW for a one-meter free-space link. In basis-randomization Protocol \#2 (blue dots), the sifted key rate increases from 0.148 Mbps to 1.52 Mbps over the same pump-power range, yielding a slope of 0.11 Mbps/mW.} By contrast, the conventional scheme (black dots) exhibits only a modest increase from 0.045 Mbps to 0.47 Mbps, with a slope of 0.035 Mbps/mW. Such a superior performance of our current BBM92 with post detected basis randomization protocols ($\#$ 1 and 2) in terms of higher sifted key rate with respect to the conventional BBM92 scheme can be understood as follows. Let us consider one of the derived sources, EPS1. In the absence of a beam splitter, the coincidence rate between Alice and Bob in the same basis (e.g., Z) is $N_{CC}$. Introducing an ideal 50:50 beam splitter at each photon path of Alice and Bob reduces the useful coincidences to the transmitted–transmitted and reflected–reflected outputs, leading to a 50\% loss \cite{neumann2021model} and a key rate of $N_{CC}/2$. \GKS{In Protocol \#1, the measurement basis for coincident events is assigned in a correlated manner, thereby eliminating the conventional loss arising from basis mismatch. In contrast, Protocol \#2 employs independent basis assignment at Alice and Bob, leading to a 50\% loss of coincident events due to basis mismatch, similar to the standard BBM92 protocol. Moreover, our scheme avoids beam splitters and utilizes two entangled photon sources, EPS 1 and EPS 2. Consequently, Protocol \#1 and Protocol \#2 yield sifted key rates of $2N_{CC}$ and $N_{CC}$, corresponding to fourfold and twofold enhancements, respectively, compared to the conventional BBM92 scheme.}

Experimentally, we observe more than sixfold and three fold enhancement (Fig. \ref{Figure3}(a)) in the sifted key rate for Protocol $\#$1 and Protocol $\#$2 with respect to conventional BBM92 protocol under similar experimental conditions, which we attribute to non-ideal splitting ratios, inherent bias, and losses in practical beam splitters.

Further, we quantify the QBER using the following equations,

\begin{align}
R_{sif} = & CC_{HH} + CC_{VV} + CC_{DD} + CC_{AA} \notag\\
& + CC_{HV} + CC_{VH} + CC_{DA} + CC_{AD}
\label{Eq1}
\end{align}
and 
\begin{equation}
QBER = \frac{CC_{HV} + CC_{VH} + CC_{DA} + CC_{AD}}{R_{sif} }
\label{Eq2}
\end{equation}

where $CC_{HH}, \dots, CC_{AA}$ and $CC_{HV}, \dots, CC_{AD}$ denote coincidence counts in the correct and incorrect bases, respectively. As shown in Fig. \ref{Figure3}(b), the QBER in our scheme for Protocol \#1 (red dots) increases from 1.03\% to 6.44\% at a slope of 0.45\%/mW, and for Protocol \#2 (blue dots) increases from 1.43\% to 9.28\% at a slope of 0.65\%/mW, whereas the conventional approach (black dots) rises from 1.8\% to 8.04\% at 0.55\%/mW. These results confirm that the present BBM92 scheme with new basis randomization protocols, which exploits the intrinsic spatial randomness of SPDC photons, outperforms the conventional beam-splitter–based scheme by simultaneously achieving higher key rates and lower QBER.

\GKS{We computed the coincidence counts between Alice and Bob for both the present scheme and the conventional BBM92 scheme at a fixed pump power of 3 mW. The distribution of coincidence events across the $Z$ and $X$ bases provides a measure of the intrinsic randomness in linear and rectilinear basis encoding. Ideally, this distribution should approach a 1:1 ratio. For the conventional BBM92 scheme, the numbers of encoded bits in the H/V and D/A bases are 74.9 kHz and 58.42 kHz, respectively, yielding a ratio of 1.3:1 and indicating deviations from the ideal case due to beam-splitter imperfections. In contrast, for Protocol \#1 the numbers of encoded bits in the H/V and D/A bases are 436.5 kHz and 434 kHz, while for Protocol \#2 they are 220 kHz and 220 kHz, respectively, resulting in a nearly ideal 1:1 ratio. These results confirm that the spatially random distribution of SPDC photons enables unbiased, quantum-mechanical encoding, consistent with our previous report~\cite{nai2024device}.}

To provide a clear comparison between the present BBM92 scheme and the conventional beam-splitter–based BBM92 scheme, we summarize the key performance parameters including QBER, sifted key rate, error-corrected key rate, privacy-amplified key rate, and the final secure key rate in Table \ref{tab:qkd_parameters}. The calculations employ standard post-processing procedures, with error correction implemented via low-density parity check (LDPC) codes and privacy amplification performed using established algorithms. As shown in Table \ref{tab:qkd_parameters}, our spatial-randomness–based scheme with both basis randomization protocols consistently outperforms the conventional protocol across all stages of key generation, demonstrating its advantage for high-rate and secure quantum communication.


\begin{table}[H]
\centering
\caption{Comparison of QKD performance metrics between the beam-splitter–based conventional BBM92 protocol and the spatial-randomness–based implementation introduced here.}
\begin{tabular}{lccc}
\hline
\textbf{Parameters} & \textbf{Conventional} & \multicolumn{2}{c}{\textbf{Present Scheme}} \\
                    & \textbf{Scheme}       & \textbf{Protocol \#1} & \textbf{Protocol \#2} \\
\hline
Encoding ratio (H/V:D/A) & 1.3:1  & 1:1  & 1:1   \\
QBER (\%)                & 8.04  & 6.44  & 9.27   \\
Sifted key rate (Mbps)   & 0.47  & 2.90  & 1.52   \\
Post EC key rate (Mbps)  & 0.45  & 2.76  & 1.44   \\
Key rate after PA (Mbps) & 0.15  & 1.02  & 0.45   \\
Secure key rate (Mbps)   & 0.15  & 1.02  & 0.45   \\
\hline
\end{tabular}
\label{tab:qkd_parameters}
\end{table}

\section{Conclusions}
We demonstrate a new experimental implementation of an efficient and scalable BBM92 QKD protocol by exploiting the intrinsic quantum-mechanical randomness of the spatial distribution of photon pairs generated through momentum conservation in the SPDC process. Using a polarization-entangled photon source with an annular spatial-ring distribution, we derive two entangled photon sources that generate photon pairs randomly and used for measurement-basis selection, thereby replacing the beam splitters employed in conventional BBM92 schemes. Unlike the conventional approach, this method avoids bit loss arising from imperfections of physical beam splitters. \GKS{However, the spatial distinguishability of the photon pairs could, in principle, introduce security vulnerabilities. We mitigate this by performing basis randomization after photon detection using two distinct protocols that exploit the intrinsic timing jitter of the detectors, thereby concealing the basis information from a potential eavesdropper. This post-detection basis assignment effectively emulates stochastic basis selection while avoiding beam-splitter induced losses and bias. The security of the scheme nevertheless relies on the assumptions of trusted detectors, intrinsic and uncontrollable timing jitter, and the operational indistinguishability of the photon sources.}\\
\indent In a proof-of-concept demonstration over a 1 m free-space link and under identical experimental conditions, the scheme achieves sifted key rates in the megabits-per-second regime, corresponding to a 6.4-fold enhancement over the conventional BBM92 protocol, consistently lower QBER ($<7\%$) across varying pump powers, and a near-ideal encoding balance between linear and rectilinear basis thereby eliminating encoding bias. Although the present spatial demultiplexing scheme requires four quantum channels, it can be adapted to fiber-based QKD systems by exploiting energy conservation to generate two wavelength-correlated entangled sources from a single SPDC process and by using wavelength demultiplexers instead of beam splitters. Spatial- or wavelength-demultiplexing-based randomness thus provides a practical pathway toward high-rate, low-error, and bias-free entanglement-based BBM92 QKD, enabling robust and scalable quantum communication networks.

\section*{AUTHOR DECLARATIONS}
\subsection*{Conflict of Interest}
The authors have no conflicts to disclose.
\subsection*{Author Contributions}
A. K. N. developed the experimental setup and performed measurements. G. P. S. participated in experiments, A. K. N., R. G., and C. M. C. participated in data analysis, numerical simulation, and data interpretation. A. K. N. and G. K. S. developed the ideas, and G. K. S. led the project. All authors participated in the discussion and contributed to the manuscript writing.

\begin{acknowledgments}
A. K. N., G. P. S., R. G., and G. K. S. acknowledge the support of the Department of Space, Govt. of India. A. K. N. acknowledges funding support for Chanakya - PhD fellowship from the National Mission on Interdisciplinary Cyber-Physical Systems of the Department of Science and Technology, Govt. of India, through the I-HUB Quantum Technology Foundation.  G. K. S. acknowledges the support of the Department of Science and Technology, Govt of India, through the Technology Development Program (Project DST/TDT/TDP-03/2022).  
\end{acknowledgments}
\section*{DATA AVAILABILITY}
The data that support the findings of this study are available from the corresponding author upon reasonable request.

\bibliography{Bib}
\end{document}